\documentclass[12pt]{article}
\usepackage[dvips]{graphicx}
\usepackage{amssymb}

\newcommand{\eq}{\begin{equation}}
\newcommand{\en}{\end{equation}}
\newcommand{\be}{\begin{equation}}
\newcommand{\ee}{\end{equation}}
\newcommand{\eqa}{\begin{eqnarray}}
\newcommand{\ena}{\end{eqnarray}}
\newcommand{\ba}{\begin{eqnarray}}
\newcommand{\ea}{\end{eqnarray}}

\newcommand{\ZZ}{\hbox{{\rm Z{\hbox to 3pt{\hss\rm Z}}}}}

\begin{document}
\begin{titlepage}
\vskip0.5cm
\vskip0.5cm
\begin{center}
{\Large\bf
The Binder Cumulant at the Kosterlitz-Thouless Transition
}
\end{center}

\centerline{
Martin Hasenbusch
}
\vskip 0.4cm
\centerline{\sl 
Institut f\"ur Theoretische Physik, Universit\"at Leipzig, 
}
\centerline{\sl Postfach 100 920, D-04009 Leipzig, Germany}
\vskip 0.3cm
\centerline{e-mail: Martin.Hasenbusch@itp.uni-leipzig.de}
\vskip 0.4cm
\begin{abstract}
We study the behaviour of the Binder cumulant on finite square lattices
at the Kosterlitz-Thouless phase transition. We determine the fixed point value
of the Binder cumulant and the coefficient of the leading logarithmic 
correction. These calculations are supplemented with Monte Carlo simulations
of the classical XY (plane rotator) model, the Villain model
and the dual of the absolute value solid-on-solid 
model. Using the single cluster algorithm, we simulate lattices up to 
$L=4096$. For the lattice sizes reached, subleading corrections 
are needed to fit the data for the Binder cumulant. 
We demonstrate that the combined analysis of the Binder cumulant 
and the second moment correlation length over the lattice size
allows for an
accurate determination of the Kosterlitz-Thouless 
transition temperature on relatively small lattices. We test the 
new method at the example of the 2-component $\phi^4$ model on 
the lattice.
\end{abstract}

\end{titlepage}

\section{Introduction}
Two dimensional systems with short range interactions and $O(2)$ symmetry
undergo a Kosterlitz-Thouless (KT) phase transition  \cite{KT}. 
This phase transition is of particular interest
because of its peculiar nature and the large number of effectively 
two dimensional systems that are supposed to undergo such a transition.

Following the theorem of Mermin and Wagner \cite{MW} the magnetisation of
two dimensional systems with short range interactions and a continuous 
symmetry vanishes at any finite temperature.
Kosterlitz and Thouless \cite{KT} have argued that nevertheless two dimensional
systems with $O(2)$ symmetry undergo a phase transition at a finite temperature.
In the low temperature phase, 
the decay of the correlation function 
follows a power law.
 At a sufficiently high temperature, pairs of vortices unbind and
disorder the system, resulting in a finite correlation length $\xi$; i.e.
the correlation function decays exponentially.
Starting from the seminal work of Kosterlitz and Thouless a rather 
solid theoretical understanding of the transition has been established.
See e.g. refs. \cite{Jo77,AmGoGr80}.  Furthermore, 
there are exactly solved models
\cite{lieb,wu,baxter,beijeren77} that display the behaviour predicted by  
KT-theory.

This excellent theoretical understanding is contrasted by the fact that Monte 
Carlo studies of the KT-transition are notoriously difficult.
These difficulties are related with logarithmic corrections that are
present in the neighbourhood of the transition. 

In Monte Carlo studies of critical phenomena, renormalization group
invariant quantities, which are also called phenomenological couplings,
are very useful tools.  
In finite size scaling (FSS), they allow to locate the transition point,
to determine the nature of the transition and, in the case of a second 
order transition, to determine the critical exponent $\nu$ of the correlation 
length. 
The prototype of such a quantity is the so called Binder cumulant
\begin{equation}
 U = \frac{\langle (\vec{m}^2)^2 \rangle}{\langle \vec{m}^2 \rangle^2} \;\;,
\end{equation}
where $\vec{m}$ is the magnetisation of the system
\footnote{The standard convention is $U =1-\frac{1}{3} 
\frac{\langle (\vec{m}^2)^2 \rangle}{\langle \vec{m}^2 \rangle^2} $.}.
While the Binder cumulant is a standard tool in the study of second 
order transitions, only a few authors have advocated its use 
in the case of a KT-transition \cite{loison,wysin}.
This is due to the fact that little is known about the behaviour of the Binder 
cumulant in the neighbourhood of the KT-transition.  The main 
purpose of the present paper is to fill this gap.

The outline of the paper is the following: First we define the models and the 
observables that we study.
Next we summarize the results from KT-theory which are relevant to our 
problem. Then we derive the fixed point value of the Binder
cumulant and the coefficient of the leading logarithmic correction. 
We discuss our Monte Carlo simulations of the XY model, the Villain model
and the dual of the absolute value solid-on-solid (ASOS) model
at the KT-transition.
The Monte Carlo results for the Binder cumulant, the second moment correlation 
length and the helicity modulus are confronted with the theoretical 
predictions. Based on this discussion, 
we propose to determine the transition temperature of a model which 
is supposed to undergo a KT-transition
by simultaneously matching the values of 
the second moment correlation length and the 
Binder cumulant with those obtained for the three models studied in this paper.
This method is tested at the example of the two-component $\phi^4$ model 
on the square lattice. Finally we give our conclusions.

\section{The models}
We study the XY model on the square lattice and two generalizations of it:
The Villain model and the dual of the absolute value solid-on-solid model.
This allows us to check the universality 
of the behaviour of the Binder cumulant and other phenomenological couplings.
The Boltzmann factor of these 
models can be written as a product of weights for pairs of nearest 
neighbour sites on the lattice. 
In all three cases we have
\begin{equation}
\label{XYboltz}
B(\{\vec{s}_x\})  = \prod_{x,\mu}  w(\theta_{x,\mu}) \;\;,
\end{equation}
where $\theta_{x,\mu}$ is the angle between nearest neighbour spins $\vec{s}_x$ 
and $\vec{s}_{x+\hat \mu}$.
The spin $\vec{s}_x$ is a unit vector with two real components, 
 $x=(x_1,x_2)$ labels the sites on the square lattice,
where $x_1 \in \{1,2,...,L_1\}$ and $x_2 \in \{1,2,...,L_2\}$
\footnote{In our simulations
we use $L_1=L_2=L$ throughout}, 
$\mu$ gives the direction on
the lattice and $\hat \mu$ is a unit-vector in the $\mu$-direction. 
We consider periodic boundary conditions in both directions.
The weight function is periodic: $w(\theta)=w(\theta + 2 n \pi)$ for any 
integer $n$. The partition function is given by the integral 
\begin{equation}
Z = \int \prod_x [\mbox{d} s_x^{(1)}  
\mbox{d} s_x^{(2)} \delta(\vec{s}_x^{\;2} -1)]
 \;  B(\{\vec{s}_x\}) \;\;.
\end{equation}

Note that two dimensional XY-models on square lattices with Boltzmann 
factors given by eq.~(\ref{XYboltz}) can be exactly mapped onto so called 
solid-on-solid (SOS) models. For a detailed discussion see ref. \cite{savit}.
The variables $h_x$ of an SOS model are integers living on the sites $x$ of
a square lattice.
The Boltzmann factor of these models can be written as a product
over nearest neighbour pairs
\begin{equation}
\label{SOSboltz}
\tilde B(\{\vec{h}_x\})  = \prod_{x,\mu}  \tilde w(|h_x-h_{x+\hat \mu}|) \;\;,
\end{equation}
The relation between the weight function of an SOS model and its  dual
XY model is given by
\begin{equation}
 w(\theta) = 
 \sum_{n=-\infty}^{\infty} \tilde w(n) \cos(n \theta)
\end{equation}
and 
\begin{equation}
\tilde w(n) = \frac{1}{2 \pi} \int_{-\pi}^{\pi} \mbox{d} \theta \; w(\theta)
                                         \cos(n \theta) \;.
\end{equation}

The weight of the standard XY model (or plane rotator model) is given by
\begin{equation}
w_{XY}(\theta_{x,\mu}) = \exp(\beta \cos(\theta_{x,\mu})) = 
                 \exp(\beta \; \vec{s} \cdot \vec{s}_{x+\hat \mu}) \; ,
\end{equation}
where $\beta$ is the inverse temperature.  In the case of the other two 
models it is simpler to give the weights for the SOS representation.
The Villain model is dual to the discrete Gaussian solid-on-solid
(DGSOS) model. 
Its weight is given by
\begin{equation}
\tilde w_{DGSOS}(n) = \exp(-\tilde \beta n^2 ) \;.
\end{equation}
It follows
\begin{equation}
w_{Villain}(\theta) = \sum_{n=-\infty}^{\infty}  
\exp(-\tilde \beta n^2 +i n \theta) = c
\sum_{m=-\infty}^{\infty} 
\exp\left(-\frac{\beta}{2} (\theta -2 \pi m)^2\right) \;\;, 
\end{equation}
where $\beta=1/(2 \tilde \beta)$ and $c = \sqrt{2 \pi \beta}$.
For a proof of this equation see e.g. appendix A.1 of ref. \cite{korzek}.

The weight of the absolute value solid-on-solid (ASOS) model is given by
\begin{equation}
\tilde w_{ASOS}(n) = \exp(-\tilde \beta |n| ) \;\;.
\end{equation}

The KT-temperatures of these models have been accurately determined 
\cite{HaMaPi94,HaPi97} by an 
RG mapping of these models with the exactly solved body centred cubic 
solid-on-solid (BCSOS) model \cite{lieb,wu,baxter,beijeren77}: 

\begin{eqnarray}
\beta_{KT}^{XY}&=&1.1199(1) \;\;\;\;\;\;\;\;\;\;\;\;\;   b_m^{XY} = 0.93(1) \;, \nonumber \\
\beta_{KT}^{Villain}&=&0.75154(23)   \;\;\;\;\;\;    b_m^{Villain} =0.32(1) \;, \nonumber \\
\tilde \beta_{KT}^{ASOS}&=&0.80608(2)  \;\;\;\;\;\;\;\;\;b_m^{ASOS} =2.78(3) \;.
\label{ourresults}
\end{eqnarray}
In addition to the (inverse) temperature of the KT-transition 
we give the scale factor
$b_m$ that is needed to match the given model with the BCSOS model.  
E.g. results for the ASOS model at the KT-transition obtained at the lattice 
size $L_{ASOS}$ match 
with the BCSOS model at the transition at the lattice size 
$L_{BCSOS}=L_{ASOS}/b_m^{ASOS}$.

\section{The observables}
In this section we shall summarize the definitions of the observables 
that we have measured in our simulations.
The total magnetisation is defined by
\begin{equation}
 \vec{m} =  \sum_x \vec{s}_x \;\;.
\end{equation}
The magnetic susceptibility is then given  by
\begin{equation}
\label{chi}
 \chi =  \frac{1}{L^2} \langle \vec{m}^2 \rangle \;\;,
\end{equation}
where $ \langle \ldots \rangle$ denotes the expectation value with respect to
the Boltzmann factors defined in the previous section.
For completeness we repeat the definition of the Binder cumulant 
\begin{equation}
 U = \frac{\langle (\vec{m}^2)^2 \rangle}{\langle \vec{m}^2 \rangle^2} \;.
\end{equation}

The second moment correlation length on a lattice of the 
size $L^2$  is defined by
\begin{equation}
\label{second}
\xi_{2nd}=\frac{1}{2 \sin(\pi/L)} \left(\frac{\chi}{F}-1\right)^{1/2} \;,
\end{equation}
where  $\chi$ is the magnetic susceptibility as defined above
and
\begin{equation}
 F =\frac{1}{L^2} 
\sum_{x,y} \langle \vec{s}_x \vec{s}_y \rangle  \cos(2 \pi (y_1-x_1)/L) 
\end{equation}
is the Fourier transform of the correlation function at the smallest 
non-vanishing momentum. In our simulations we have measured $F$ for both 
directions of the lattice to reduce the statistical error.

The helicity modulus $\Upsilon$ gives the reaction of the system under
a torsion \cite{helidef}.
To define the helicity modulus we consider a system, where
rotated boundary conditions in one direction are introduced:
For pairs $x,x+\hat \mu$ of nearest neighbour sites on the lattice with 
$x_1=L_1$ and $\mu=1$ the weight  $ w(\theta_{x,\mu})$
is replaced by $ w(\theta_{x,\mu} + \alpha)$.
The helicity modulus  
is then defined by the second derivative of the
free energy with respect to $\alpha$ at $\alpha=0$
\begin{equation}
\label{ups_inf}
\Upsilon = -  \frac{L_1}{L_2}
\left .
\frac{\partial^2 \ln Z(\alpha)}{\partial \alpha^2} \right |_{\alpha=0} \;\;.
\end{equation}
Note that we have skipped a factor one over temperature in our definition 
of the helicity modulus to obtain a dimensionless quantity. In the case of the 
standard XY model
it is easy to write the helicity modulus as an observable of the 
system at $\alpha=0$ \cite{heliform}. For $L_1=L_2=L$ we get
\begin{equation}
\label{heliexpr}
\Upsilon = \frac{\beta}{L^2} 
\left \langle \vec{s}_{x \phantom{\hat 1} } \vec{s}_{x+\hat 1} \right \rangle
-  \frac{\beta^2}{L^2} \left \langle \left(
  s_{x\phantom{\hat 1}}^{(1)} s_{x+\hat 1}^{(2)} 
- s_{x\phantom{\hat 1}}^{(2)} s_{x+\hat 1}^{(1)} 
\right)^2 \right \rangle \;\;.
\end{equation}
Note that this equation is not valid for the other two models that we have
studied.

\section{KT-theory}
\label{KTtheory}
At low temperatures, fluctuations are suppressed and we might expand the 
weight as
\begin{equation}
 -\ln(w(\theta))  =  const - \frac{\beta_{SW}}{2} \theta^2 + \ldots 
\end{equation}
Note that for the models discussed above, $w(\theta)$ is an even function 
that assumes its maximum  at $\theta=0$. 
Using this approximation we arrive at the exactly solvable Gaussian model
(or free field theory in the language of high energy physics):
\begin{equation}
\label{zgauss}
Z_{SW} = \int \prod_x [\mbox{d} \theta_x]  \delta\left(\sum_x \theta_x\right)
\exp(-H_{SW})
\end{equation} 
with
\begin{equation}
\label{HSW}
H_{SW} = \frac{\beta_{SW}}{2}  \sum_{x,\mu}   (\theta_x - \theta_{x+\hat \mu})^2
\;\;.
\end{equation}
Note that the $\delta$-function in eq.~(\ref{zgauss}) is needed to render the
integral finite.
For the solution of this model see e.g. Appendix A of ref. \cite{Jo77}
or textbooks; e.g. ref. \cite{bellac}.

In the spin-wave approximation, vortices that drive the KT phase 
transition are absent.  A careful analysis shows that they are, in an RG-sense,
irrelevant for $\beta_{SW} \ge \frac{2}{\pi}$ \cite{Jo77,AmGoGr80}.  I.e. 
the large distance  behaviour of systems at the KT-transition 
is given by the spin-wave model at $\beta_{SW} = \frac{2}{\pi}$. 

For the discussion of the RG-flow in the neighbourhood of  the KT-transition
it is convenient to  define
$ x = \pi  \beta_{SW} -2$.
At the KT-transition, $x$ behaves as  \cite{Jo77,AmGoGr80}
\begin{equation}
\label{xtransition}
 x(l) = \frac{1}{\ln l + C} + \ldots \;,
\end{equation}
where $C$ is an integration constant 
and $l$ is a length scale. In the case of finite size scaling,
we identify the lattice size $L$ with this scale.

Leading corrections to the asymptotic behaviour  of
correlation functions at the KT-transition can be obtained by computing
the correlation function in the spin-wave approximation for $\beta_{SW}$ given 
by eq.~(\ref{xtransition}).

Along these lines we obtained 
\cite{myrecent}
\begin{equation}
\label{centralheli}
\Upsilon_{L^2,transition} = 0.63650817819... + \frac{0.318899454...}
{\ln L + C} + \;... 
\end{equation}
for the helicity modulus.
Note that in the literature 
(see e.g. eq.~(4) of ref. \cite{WeMi88}) often
$\Upsilon=\frac{2}{\pi} + \frac{1}{\pi} \frac{1}{ (\ln L + C)} \ldots $
is given. The small difference is due to the fact that in ref. \cite{myrecent}
contributions from configurations with non-zero winding number are 
taken into account.
For the second moment correlation length over the lattice size 
we obtained  \cite{myrecent}
\begin{equation}
\label{xiexact}
\left . \frac{\xi_{2nd}}{L} \right |_{ L^2,transition} = 0.7506912... + \frac{0.212430...}{\ln L + C}+ \;... \;\;.
\end{equation}
The subscript refers to the fact that the result holds for a lattice of the size
$L_1=L_2=L$ with periodic boundary conditions at the KT-transition.
Furthermore the couplings between the spins
have to be the same in both directions as it is the case here.

Computing four-point functions 
$\langle (\vec{s}_x \cdot \vec{s}_y) (\vec{s}_u \cdot \vec{s}_v) \rangle$
for all $y-x$, $u-x$ and $v-x$
from the propagator of the Gaussian model requires a triple sum over all 
points of the lattice. Therefore we could not reach sufficiently large 
lattice sizes this way. To avoid this problem, we performed Monte 
Carlo simulations of the Gaussian model instead. 

\subsection{Monte Carlo Simulation of the Gaussian Model}
First note that $\beta_{SW}$ in eq.~(\ref{HSW}) can be absorbed into the 
field variable by
$\phi_x = \beta_{SW}^{1/2} \theta_x$. In terms of the new field variable
the Hamiltonian becomes
\begin{equation}
\label{HSW2}
 H_{SW}= \frac{1}{2} \sum_{x,\mu} (\phi_x - \phi_{x+\hat \mu})^2 \;\;.
\end{equation}
The spin configurations are given by
\begin{equation}
\label{vectgauss}
 \vec{s}_x(n_1,n_2) = \left(\cos(\beta_{SW}^{-1/2}  \phi_x + \psi_x(n_1,n_2)), 
      \sin(\beta_{SW}^{-1/2} \phi_x + \psi_x(n_1,n_2)) \right) \;,
\end{equation}
where 
$\psi_x(n_1,n_2)=2 \pi (n_1 x_1 + n_2 x_2)/L$ takes into account windings of the 
spin configuration, where $n_1$ and $n_2$ are integers. As discussed in section
3.2 of ref. \cite{myrecent},  expectation values of an observable $A$ 
in the spin wave approximation on a lattice with periodic boundary 
conditions are given by
\begin{equation}
\label{swobs}
 \langle A(\{\vec{s}\})   \rangle_{SW}  
 = \frac{\sum_{n_1,n_2} W(n_1,n_2) \langle A(\{\vec{s}\}_{n_1,n_2})  \rangle}
  {\sum_{n_1,n_2}  W(n_1,n_2)} \;\; ,
\end{equation}
where
\begin{equation}
 \langle A(\{\vec{s}\}_{n_1,n_2})  \rangle
=\frac{\int \prod_x [\mbox{d} \phi_x] \delta(\sum_x \phi_x)
  \exp(-H_{SW}(\{ \phi \})) A(\{\vec{s}\}_{n_1,n_2})}
    {  \int \prod_x [\mbox{d} \phi_x] \delta(\sum_x \phi_x) 
    \exp(-H_{SW}(\{ \phi \} ))}  \;\;.
 \end{equation}
The weights for the different winding numbers are given by
\begin{equation}
 W(n_1,n_2) = \exp\left(- 2 \pi^2 \beta_{SW}  [n_1^2  + n_2^2]  \right) \;\;.
\end{equation}
Note that these weights do not depend on $L$ and therefore also non-zero
winding numbers contribute to the asymptotic behaviour.
In our simulations we have only taken into account the winding numbers
$(n_1,n_2)=$ $(-1,0), (1,0), (0,-1)$ and $(0,1)$. The weight for these winding 
numbers is $0.000003487\ldots$ at $\beta_{SW}=2/\pi$.  This weight is rather 
small. However it turns out that at the level of accuracy that we have reached
these windings have to be taken into account. Higher winding numbers can be 
safely ignored.

In momentum space, the degrees of freedom of the Gaussian model decouple.
Therefore one can directly generate a field configuration, using e.g. 
the Box-Muller algorithm, in momentum space with the correct probability density.
Then one can perform a Fourier transformation to obtain the
configuration in real space.

Since we had the program code available, we used a different 
approach: We generated the configurations directly in real space, using 
a mixture of the local Metropolis, the overrelaxation and a single 
cluster version \cite{wolff} of the valley to mountain 
reflection (VMR) algorithm \cite{vmr}. Due to the use of the cluster 
algorithm, there should be no slowing down and therefore our choice should
have a similar performance as the one sketched above.

In an elementary step of the Metropolis algorithm we propose to change the 
field at the site $x$ as 
\begin{equation}    
 \phi_x ' = \phi_x + 2 (r -1/2)  \;,
\end{equation}
were $r$ is a random number that is uniformly distributed in $[0,1]$. 
The proposal is accepted with the probability 
$A=\mbox{min}[1,\exp(-\Delta H)]$. 
An elementary step of the overrelaxation algorithm is given by
\begin{equation}
 \phi_x ' = 2 \frac{\sum_{y.nn.x} \phi_y}{4} -  \phi_x   \;,
\end{equation}
where $y.nn.x$ indicates that $y$ is a nearest neighbour of $x$.
Note that this update does not change the value of the Hamiltonian. 
In both cases, the Metropolis update and the overrelaxation update, we go 
through the lattice in lexicographic order. Going through the lattice once
with the local update is called sweep.

The variant of the VMR algorithm that we have used here is given by the 
following steps:

\begin {itemize}
\item
Chose randomly a site $x$ of the lattice. The reference hight for the 
update is then given by $h_0= \phi_x$. All fields $\phi_z$ that reside on 
sites $z$ that belong to the cluster  are updated as
\begin{equation}
 \phi_z ' = 2 h_0 -  \phi_z   \;.
\end{equation}
I.e. the field is reflected at $h_0$.

\item
Chose randomly a site $y$ of the lattice as the starting point of the 
single cluster. The cluster consists of all sites $z$ that are connected
with $y$ by a chain of frozen links. The links that are not frozen are called
deleted.
The probability to delete a link $v,\mu$ is given by 
\begin{equation}
 p_d(v,\mu) = 
  \mbox{min}[1,\exp(-2 (\phi_v- h_0) (\phi_{v+\hat \mu} - h_0) ) ] \;\;.
\end{equation}
Note that in this equation both $\phi_v$ and $\phi_{v+\hat \mu}$ are
taken before the update.
\end{itemize}

We performed the updates in the following sequence: One Metropolis sweep, 
one overrelaxation sweep and finally a VMR single cluster update. 
The average size of the single cluster is about 1/3 of the lattice, independent
of the lattice size.  Integrated autocorrelation times of the magnetic 
susceptibility~(\ref{chi},\ref{swobs})
at $\beta_{SW}=2/\pi$ are about 2.8 for $L=32$ and increase to about 5 for 
$L=512$, where the time unit is one update sequence as specified above.

In our production runs, the measurements are separated by 5 such sequences.
We performed  $20$, $20$, $12$, $5$ and $2.2$  $\times 10^7$ 
measurements for 
$L=32$, $64$, $128$, $256$ and $512$, respectively.  In total, these 
simulations took about 10 month of CPU time on a 3 GHz Pentium 4 CPU. 

In order to compute $\lim_{L\rightarrow \infty} U_4$ at the KT-transition, 
we have
set $\beta_{SW}=\frac{2}{\pi}$ in eq.~(\ref{vectgauss}). In addition we have
used $\beta_{SW} =\left(\sqrt{\frac{2}{\pi}}+0.0001 \right)^2$ and 
$\beta_{SW} =\left(\sqrt{\frac{2}{\pi}}-0.0001 \right)^2$
to compute the derivative of 
$\lim_{L\rightarrow \infty} U_4$ with respect to $\beta_{SW}$ at 
$\beta_{SW}=\frac{2}{\pi}$. Note that we have used the same set of 
$\phi$ configurations for the three values of $\beta_{SW}$. 
Our results are summarized in table \ref{binderGAUSS}.

A brief remark on non-zero winding numbers:
Taking only zero winding
configurations, $U_4$ is about $0.000013$ smaller than with the 
non-zero winding numbers taken into account. I.e. the contribution of 
non-zero winding numbers is a little larger than our statistical error.

\begin{table}
\caption{\sl \label{binderGAUSS}
Results for the Binder cumulant $U_4$ in the  spin wave approximation at 
$\beta_{SW} =2/\pi$ and its derivative with respect to $\beta_{SW}$. In the 
last row we give the result for the limit $L \rightarrow \infty$.
}
\begin{center}
\begin{tabular}{|r|r|r|}
\hline
  $L$& $U_4$ \phantom{xxx} &$\mbox{d} U_4/\mbox{d}{\beta_{SW}}$ \\
\hline
 32& 1.018554(2) & --0.057529(7)\phantom{0}\\
 64& 1.018298(2) & --0.056679(7)\phantom{0}\\
128& 1.018217(3) & --0.056394(9)\phantom{0}\\
256& 1.018200(5) & --0.056332(15)\\
512& 1.018199(8) & --0.056297(25)\\
\hline
$\infty$ &1.018192(6) & 
  --0.056303(16)\\
\hline
\end{tabular}
\end{center}
\end{table}

In order to obtain a result for the limit $L \rightarrow \infty$ we 
have fitted the results to the ansatz $X(L) = X(\infty) + c L^{-2}$, 
where $X$ is either the Binder cumulant or its derivative.
Our final results are taken from the fit that includes the lattice sizes 
$L=128$, $256$ and $512$. These results are consistent within error bars
with those obtained from $L=64$ and $128$.  
Hence the systematic error due to higher order corrections should be quite small.

Plugging the result for the derivative of $U_4$ into eq.~(\ref{xtransition})
we arrive at
\begin{equation}
\label{binderleading}
U_{4,L^2,transition} = 1.018192(6) -  \frac{0.017922(5)}{\ln L + C}  + ... \;,
\end{equation}
where we should note again that the result only holds for a lattice with
$L_1=L_2=L$ with periodic boundary conditions at the KT-transition. 
Furthermore the couplings between the spins
have to be the same in both directions as it is the case here.

\section{Monte Carlo Simulations}
In this section we discuss the details of our Monte Carlo simulations
of the XY, the Villain and the dual of the ASOS model. 

\subsection{Details of the Simulations}
We have simulated the XY model, the Villain and the dual of the ASOS model 
at the best estimates of the KT-temperature given in 
eq.~(\ref{ourresults}).  In our simulations we have used the single 
cluster algorithm \cite{wolff}. Let us briefly summarize the steps of the 
cluster update:
\begin{itemize}
\item
Chose a direction:
\begin{equation}
 \vec{d} = (\cos(2 \pi r), \sin(2 \pi r)) \; ,
\end{equation}
where $r$ is a random number which is uniformly distributed in 
$[0,1]$ \footnote{In analogy with the VMR algorithm for the Gaussian 
model, one could chose randomly some site $z$ of the lattice and then
chose $\vec{d} = \vec{s}_z$. We did note test this alternative.}.
All spins $\vec{s}_x$ that reside on
sites $x$ that belong to the cluster will be updated as
\begin{equation}
 \vec{s}_x^{\;\;'} = \vec{s}_x 
- 2 (\vec{d} \cdot \vec{s}_x) \vec{d} \;\;,
\end{equation}
\item
Chose randomly a site $y$ as starting point of the single cluster.
All sites $x$ belong to the cluster that are connected by a frozen
chain of links with the site $y$. A link that is not frozen is called
deleted. The probability to delete a link $x,\mu$ is given by
\begin{equation}
p_d(x,\mu) = \mbox{min}\left[1,
\frac{
w\left(\vec{s}_x \cdot \vec{s}_{x+\mu}
 -2 (\vec{d} \cdot \vec{s}_x) (\vec{d} \cdot \vec{s}_{x+\mu})\right)}
    {w(\vec{s}_x \cdot \vec{s}_{x+\mu})}\right] \;\;.
\end{equation}
Note that in this equation both $\vec{s}_x$ and $\vec{s}_{x+\mu}$ are
taken before the update.
\end{itemize}

Our first set of simulations of the XY-model was performed with the 
single cluster algorithm alone. In these simulations we have used the 
G05CAF random number generator of the NAG-library.  We simulated 
the lattice sizes 
$L=$ $16$, $32$, $64$, $128$, $256$, $512$, $1024$, $2048$ and $4096$. 
For $L\le 1024$ we  performed $5 \times 10^6$ measurements. For $L=2048$ and
$L=4096$ we performed $2.5 \times 10^6$ and $1 \times 10^6$ measurements, 
respectively. In all these cases, we performed 10 single cluster updates
for one measurement.
In total, these simulations
took a bit more than 7 month of CPU time on a 3 GHz Pentium 4 CPU.

Later we added simulations for $L=12$, $24$, $48$, $96$, $192$, $384$
and $768$.  
As random number generator we have used here
the  SIMD-oriented Fast Mersenne Twister (SFMT)
\cite{twister} generator.  In particular we use the function
{\sl genrand$\_$res53()} that produces double precision output.

In these simulations we performed overrelaxation updates 
in addition to the cluster updates.
An overrelaxation update of the 
spin at the site $x$ is given by
\begin{equation}
 \vec{s}_x^{\;\;'} = \vec{s}_x 
- 2 \frac{\vec{S}_x \cdot \vec{s}_x}{\vec{S}_x^2} \vec{S}_x \;\;,
\end{equation}
where 
\begin{equation}
 \vec{S}_x = \sum_{y.nn.x}  \vec{s}_y \;\;,
\end{equation}
where $y.nn.x$ means that $y$ is a nearest neighbour of $x$. It is easy
to check that this update keeps the value of the Hamiltonian constant.

For $L \le 192$ we performed $10^7$ measurements. For each measurement we
performed 
10 single cluster updates and 10 overrelaxation sweeps.
For $L =384$ and $768$ we performed $9 \times 10^6$ measurements. In these
two cases we performed 10 single cluster updates and 5 overrelaxation sweeps
for each measurement.
In total, these simulations
took a bit more than 4 month of CPU time on a 3 GHz Pentium 4 CPU.

In the case of the Villain model and the dual of the ASOS model, we 
have implemented the weight function $w(\vec{s}_x \cdot \vec{s}_{x+\hat \mu})$  
for the links as a table  
 with 100001 entries for the 
arguments $-1,-0.99998,-0.99996,...,1$.  Then, for a given value of the 
argument, we linearly interpolate between the two closest values
contained in the table.  This way, the 
maximal error in ratios of the weight for links is about $10^{-9}$ and 
$1.4 \times 10^{-9}$ for the Villain model and the dual of the ASOS model,
respectively. For  all simulations of the Villain model and the dual of the 
ASOS model we have used the SFMT random number generator.

The Villain model was simulated on lattices of the size $L=16$,$32$,$64$,$128$,
$256$,$512$,$1024$ and $2048$. For $L\le 128$ we performed $10^7$ measurement
for each lattice size. 
The statistics for the larger lattice sizes is  about  $8.2 \times 10^6$, 
$5.1 \times 10^6$, $3.1 \times 10^6$ and $1.4 \times 10^6$ measurements for 
$L=256$, $512$, $1024$ and $2048$, respectively. 
In all cases, we performed
10 single cluster updates per measurement.
In total, the 
simulations of the Villain model 
took about 3 month of CPU time on a 3 GHz Pentium 4 CPU.

In the case of the dual of the ASOS model we have simulated lattices of 
the size $L=$
16,24,32,48,64,96,128,192,256,384,512,768 and 1024.  Also here  we 
performed 10 single cluster updates per measurement. 
The number of measurements was $2 \times 10^7$ for all lattice sizes up to 
$L=512$. We performed $1.27 \times 10^7$ and $1.1 \times 10^7$ measurements
for $L=768$ and $L=1024$, respectively.
In total, the  simulations of the dual of the ASOS model
took a little less than 10 month of CPU time on a 3 GHz Pentium 4 CPU.

For all cases, 
the integrated autocorrelation times for e.g. the magnetic susceptibility 
are close to one in units of measurements. We have discarded at least the 
first 1000 measurements for equilibration.  
Given the small autocorrelation times this is by far sufficient.

\subsection{Analysis of the Data}
First we have analysed the data for the helicity modulus of the XY model.
We compared these data with the behaviour given by eq.~(\ref{centralheli}).
To this end, we have approximated the helicity modulus in the neighbourhood
of the simulation point by the Taylor expansion around the simulation point
to linear order.  I.e. we have used the ansatz
\begin{equation}
\label{helifits}
 \Upsilon(1.1199) + \left . \frac{\mbox{d} \Upsilon}{\mbox{d} \beta} 
                    \right |_{\beta=1.1199} \Delta \beta = 
		    0.63650817819 + \frac{0.318899454}
		    {\ln L + C} 
\end{equation}
with the two parameters $\Delta \beta$ and $C$ to fit our data for the 
helicity modulus.  The results  of these fits 
are summarized in table \ref{helifit}. 
In these fits we have used all data starting from some minimal lattice size
$L_{min}$ up to the maximal size available.
In order to get a $\chi^2/$d.o.f.
close to one, $L_{min} = 384$ is needed.
The result for $\Delta \beta$ is decreasing with increasing $L_{min}$. 
 Starting from 
$L_{min}=384$,  the result for $\Delta \beta$ is consistent within error-bars 
with $\Delta \beta=0$; i.e. our previous estimate for $\beta_{KT}$. Since 
corrections to the ansatz~(\ref{helifits}) are expected to decay slowly 
with increasing 
lattice size, it is difficult to give a reliable estimate of systematic
errors.
Therefore we abstain from
giving a final estimate for $\beta_{KT}$ obtained from these fits.

\begin{table}
\caption{\sl \label{helifit}
Fits of the helicity modulus $\Upsilon$ for the XY model with the 
ansatz~(\ref{helifits}).
$L_{min}$ is the minimal lattice size that is included in the fit. 
$C$ and $\Delta \beta$  are the parameters of the fit.
For a discussion see the text.
}
\begin{center}
\begin{tabular}{|r|l|c|c|}
\hline
 $L_{min}$ & \phantom{xx}  $C$ &    $\Delta \beta$  &  $\chi^2/$d.o.f. \\
\hline
  96 &   1.001(6) &   0.00049(2)  &   9.83 \\
 128 &   1.044(10)&   0.00038(3)  &   6.22 \\
 192 &   1.067(11)&   0.00033(3)  &   4.51 \\
 256 &   1.124(19)&   0.00021(4)  &   2.53 \\
 384 &   1.167(24)&   0.00013(5)  &   0.56 \\
 512 &   1.170(37)&   0.00013(7)  &   0.74 \\
 784 &   1.191(51)&   0.00009(9)  &   0.93 \\
\hline
\end{tabular}
\end{center}
\end{table}

 Next we have studied the second moment correlation length $\xi/L$ of the 
 XY-model using the ansatz 
\begin{equation}
\label{xi2fits}
 \frac{\xi(1.1199)}{L} + \frac{1}{L} \left . \frac{\mbox{d} \xi}{\mbox{d} \beta} 
                    \right |_{\beta=1.1199} \Delta \beta = 0.7506912
		     + \frac{0.212430}
		    {\ln L + C} \;.
\end{equation}
 Our results are summarized in table \ref{xifitXY}. In contrast to 
the helicity modulus, the $\chi^2$/d.o.f. is rather small already for 
$L_{min}=32$. Also the dependence of the result for $ \Delta \beta$ on $L_{min}$
is much smaller
than for the helicity modulus. Hence the amplitude of corrections to the 
ansatz should be much smaller than for the helicity modulus.
On the other hand, for a given $L_{min}$ the statistical error of 
$\Delta \beta$ obtained from $\xi/L$ is about three times larger than that from
$\Upsilon$.

Since $\Delta \beta$ as a function of $L_{min}$ behaves quite differently for
$\xi/L$ and $\Upsilon$ we might take the difference
as an estimate of the error due to the higher order contributions missing 
in the ans\"atze~(\ref{helifits},\ref{xi2fits}).
Starting from $L_{min}=384$ the result for $\Delta \beta$ from helicity modulus
is consistent within statistical errors 
with that obtained from $\xi/L$ for  $L_{min}\ge 32$. 
Hence the systematic error should not be larger than the statistical error 
for these $L_{min}$. 
Therefore we quote  $\beta_{KT} =1.1200(1)$  as final result for the inverse
of the KT-transition temperature of the XY model from the combined analysis of 
$\Upsilon$ and $\xi/L$.

\begin{table}
\caption{\sl \label{xifitXY}
Fits of the correlation length over the lattice size $\xi/L$ for the XY model
using the ansatz~(\ref{xi2fits}). 
$L_{min}$ is the minimal lattice size that is included in the fit.
$C$ and $\Delta \beta$ are the parameters of the fit.
For a discussion see the text.
}
\begin{center}
\begin{tabular}{|r|l|l|c|}
\hline
\multicolumn{1}{|c}{$L_{min}$}&
\multicolumn{1}{|c}{$C$}&
\multicolumn{1}{|c}{$\Delta \beta$}&
\multicolumn{1}{|c|}{$\chi^2/$d.o.f.}\\
 \hline
     32& 1.63(2)  &  0.00007(5) &  1.61 \\
     48& 1.64(2)  &  0.00005(6) &  1.71 \\
     64&  1.64(3) &  0.00004(7) &  1.89 \\
     96&  1.62(3) &  0.00009(7) &  1.82 \\
    128&  1.64(4) &  0.00005(9) &  2.02 \\
    192&  1.60(5) &  0.00012(10)&  1.94 \\
\hline
\end{tabular}
\end{center}
\end{table}

Next we have fitted  $\xi/L$ for the dual of the ASOS model with the analogue
of the
ansatz~(\ref{xi2fits}); I.e. $1.1199$ is replaced by $0.80608$ as argument
of $\xi$ and its derivative. 
The results are summarized in table \ref{xifitASOS}.
The $\chi^2/$d.o.f. becomes smaller than one starting from $L_{min}=48$. 
Also starting from $L_{min}=48$, the result for $\Delta \beta$ 
depends very little 
on $L_{min}$.  These fits suggest the new estimate $\tilde \beta_{KT}=0.80605(2)$.
This result is consistent with that of ref.\cite{HaPi97}.
\begin{table}
\caption{\sl \label{xifitASOS}
Analogue of table \ref{xifitXY} for the ASOS model. 
}
\begin{center}
\begin{tabular}{|r|l|l|c|}
\hline
\multicolumn{1}{|c}{$L_{min}$}&
\multicolumn{1}{|c}{$C$}&
\multicolumn{1}{|c}{$\Delta \beta$}&
\multicolumn{1}{|c|}{$\chi^2/$d.o.f.}\\
 \hline
 32 & 0.493(6)  & --0.000098(11) & 10.50 \\
 48 & 0.545(8)  & --0.000035(13) & \phantom{0}0.59 \\
 64 & 0.548(11) & --0.000031(15) & \phantom{0}0.64 \\
 96 & 0.548(16) & --0.000032(19) & \phantom{0}0.75 \\
128 & 0.543(20) & --0.000036(22) & \phantom{0}0.88 \\
\hline
\end{tabular}
\end{center}
\end{table}

Finally we have fitted the data for $\xi/L$ for the Villain model with 
the analogue of the ansatz~(\ref{xi2fits}). The
results are summarized in table \ref{xifitVillain}.  We get $\chi^2/$d.o.f.
close to one starting from $L_{min}=32$.   These fits suggest as estimate
for the inverse of the KT-transition temperature $\beta_{KT}=0.7517(2)$.
Again this result is consistent with that of ref.\cite{HaPi97}.

\begin{table}
\caption{\sl \label{xifitVillain}
Analogue of table \ref{xifitXY} for the Villain  model. 
}
\begin{center}
\begin{tabular}{|r|r|l|c|}
\hline
\multicolumn{1}{|c}{$L_{min}$}&
\multicolumn{1}{|c}{$C$}&
\multicolumn{1}{|c}{$\Delta \beta$}&
\multicolumn{1}{|c|}{$\chi^2/$d.o.f.}\\
\hline
16 &  2.78(2)  & --0.00009(8) &   2.44 \\
32 &  2.72(3)  &\phantom{--}0.00006(9)  &   1.30 \\
64 &  2.69(5)  &\phantom{--}0.00014(12) &   1.38 \\
128&  2.69(7)  &\phantom{--}0.00014(17) &   1.85 \\
\hline
\end{tabular}
\end{center}
\end{table}

Next let us discuss the results for the parameter $C$ in eq.~(\ref{xi2fits}).
As our final results we take $C_{XY}= 1.62(3)$ from  the fit with $L_{min}=96$
for the XY model,
$C_{Villain}=2.69(5)$  from  $L_{min}=64$ 
for the Villain model and $C_{ASOS}=0.54(2)$ from  $L_{min}=128$ 
for the dual of the ASOS model.
The differences of these constants for different models should be given 
by the logarithm of ratios of the scale factors $b_m$ summarized in 
eq.~(\ref{ourresults}). In particular we get
$C_{Villain} - C_{ASOS} = 2.15(5)$ to be compared
with $\log(b_{m,ASOS}/b_{m,Villain})=2.16(3)$,
$C_{Villain} - C_{XY} = 1.07(6)$ to be compared with 
$\log(b_{m,XY}/b_{m,Villain})=1.07(3)$ and finally
$C_{XY}-C_{ASOS} = 1.08(4) $ to be compared
with $\log(b_{m,ASOS}/b_{m,XY})=1.10(2)$.
I.e. the results for $C$ are fully consistent with the relative
scale factors $b_m$ obtained in ref.\cite{HaPi97}.

\begin{figure}
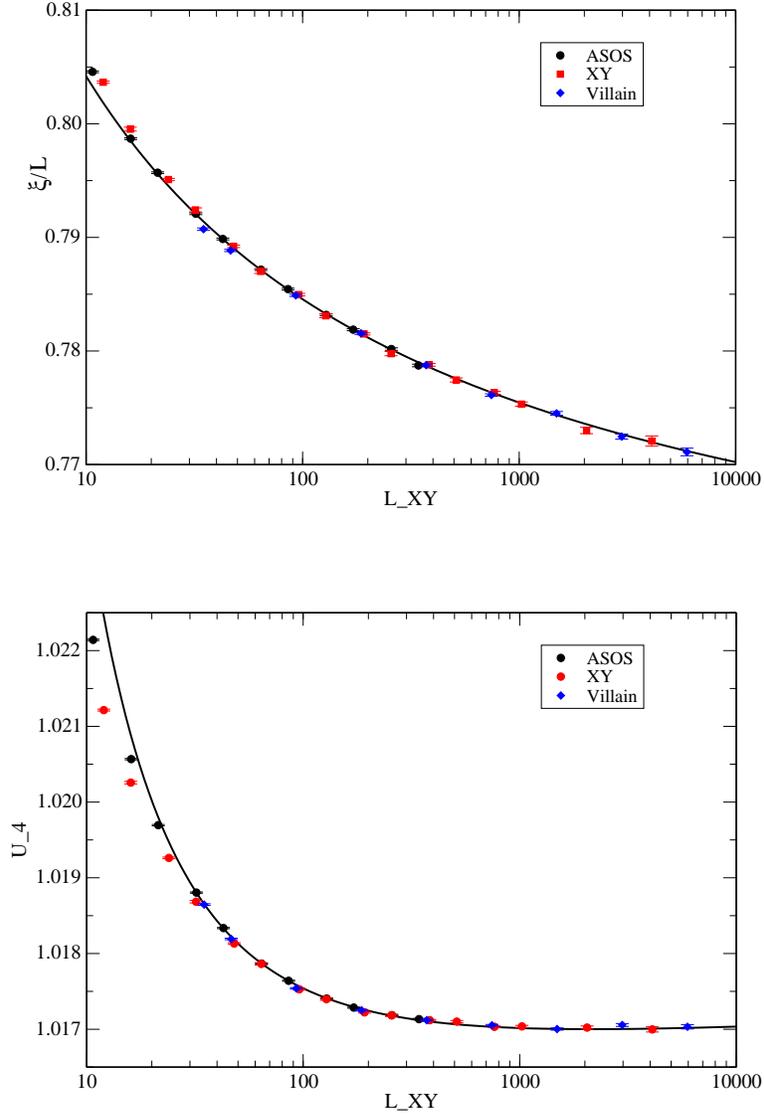

\begin{center}
\scalebox{0.40}
{
\includegraphics{xiKT.eps} 
}
\vskip0.35cm
\scalebox{0.40}
{
\includegraphics{binderKT.eps}
}
\end{center}
\caption{
\label{binderandxiL}
Data for the dual of the ASOS model, the XY model and the Villain model
at the KT transition. (See eqs.~(\ref{ourresults}))
In the upper figure we plot $\xi/L$ as a function of the lattice size
$L_{XY}$ of the XY model. Following eqs.~(\ref{ourresults}), 
the lattice size of the dual of the ASOS model and 
the Villain model are rescaled as $L_{XY}=0.3345 \times L_{ASOS}$ and 
$L_{XY}=2.906 \times L_{villain}$. 
The solid lines give $\xi/L$ and $U_4$ obtained from eqs.~(\ref{match80608}). 
      }
\end{figure}

Next we discuss the behaviour of the Binder cumulant. In fig. 
\ref{binderandxiL} we have plotted the Binder cumulant at the old estimates
of $\beta_{KT}$ summarized in eq.~(\ref{ourresults}) as a function of 
$L_{XY}=L_{model} b_{m,model}/b_{m,XY}$.
The data from the three different models fall nicely on top of 
each other, indicating a universal behaviour. 
From the leading behaviour, eq.~(\ref{binderleading}), we would expect that
the Binder cumulant is increasing with increasing lattice size $L$. However
we observe quite the opposite. For all three models, it is decreasing and 
apparently reaches a stable value $\approx 1.0170$ for large lattice sizes.

Within the KT-picture this might be explained by large  $O(1/(\ln L)^2)$ 
correction due to the effect of vortex pairs with a distance $\approx L/2$.
Configurations with such a vortex pair have a much smaller magnetisation than
those without. Therefore the appearance of such 
vortices enlarges the value of the Binder cumulant $U_4$. Since $\beta_{SW}$
is monotonically decreasing with increasing $L$, the effect seen in the Binder 
cumulant can only be explained by the effect of vortices.

Based on this observation, we have fitted the Binder cumulant at the estimate
of $\beta_{KT}$ given in eq.~(\ref{ourresults}) with the ansatz
\begin{equation}
\label{Binderansatz}
U_4(L) = 1.018192 -  \frac{0.017922}{\ln L + C} + \frac{c_2}{(\ln L + C)^2} 
\;\;,
\end{equation}
where $C$  and $c_2$ are the free parameters of the fit. Results for 
these fits for the XY model, the Villain model and the dual of the ASOS
model are given in tables \ref{binderXY}, \ref{binderVil} and \ref{binderASOS},
respectively.

$\chi^2/$d.o.f. becomes smaller close to one for $L_{min}=96$, $32$ and $96$
for the XY model, the Villain model and the dual of the ASOS model, 
respectively.  For these values of $L_{min}$, the results for $c_2$ of the 
XY model and the Villain model are fully consistent, supporting the 
universality of the value of this coefficient.  The result for the 
dual of the ASOS model is slightly smaller than that for the other 
two models. This small difference might be caused by higher order corrections
or might be due to the uncertainty in the estimate of $\beta_{KT}$.
To check the latter, we have repeated the fits for the central estimate 
for $\beta_{KT}$ 
given in eq.~(\ref{ourresults}) plus or minus the error-bar. Let us give
only a few results with $\chi^2$/d.o.f. smaller than two:
For the XY at $\beta=1.1200$ and  $L_{min}=96$ we get $C=0.006(36)$ and
$c_2=0.06764(31)$.
For the Villain model at $\beta=0.75177$ and $L_{min}=32$ we get 
$C=1.040(35)$ and $c_2=0.06699(32)$.
And finally, 
for the dual of the ASOS model at $\tilde \beta=0.80606$ and $L_{min}=96$ 
we get $C=-1.193(10)$ and $c_2=0.06730(17)$.
I.e. the small difference in the values of $c_2$ observed above can indeed 
be explained by the uncertainty of the estimate of $\beta_{KT}$.

Finally let us compare the differences in $C$ for the three models with 
the results for $b_m$ given in  eq.~(\ref{ourresults}).
Using the results obtained for
$L_{min}=96$, $32$ and $96$ for the XY-model at $\beta=1.1199$, 
the Villain model at $\beta=0.75154$ and the 
dual of the ASOS model at $\tilde \beta=0.80608$, respectively, we get: 
$C_{Villain}-C_{ASOS}=2.33(4)$, $C_{XY}-C_{ASOS}=1.25(4)$ and 
$C_{Villain}-C_{XY}=1.08(5)$. The two differences that involve the 
dual of the ASOS model are by about $0.15$ too large compared with the 
result obtained from $b_m$. The difference $C_{Villain}-C_{XY}$ is 
fully consistent with that obtained from $b_m$. 
Taking the result for the XY-model at $\beta=1.1200$,
the Villain model at $\beta=0.75177$ and the
dual of the ASOS model at $\tilde \beta=0.80606$ instead, we get
$C_{Villain}-C_{ASOS}=2.23(4)$, $C_{XY}-C_{ASOS}=1.20(4)$ and
$C_{Villain}-C_{XY}=1.03(5)$.  I.e. the deviation from the expected
result is much reduced.

\begin{table}
\caption{\sl \label{binderXY}
Fits of the Binder cumulant of the XY model
at $\beta=1.1199$ with the ansatz~(\ref{Binderansatz}).
}
\begin{center}
\begin{tabular}{|r|l|r|c|}
\hline
$L_{min}$&\phantom{000}$C$  & $c_2$\phantom{00000}  & $\chi^2/$ d.o.f.\\
\hline
 32\phantom{a}&         --0.002(14) & 0.06821(21) & 2.55 \\
 48\phantom{a}&         --0.024(18) & 0.06801(24) & 2.47 \\
 64\phantom{a}&         --0.012(27) & 0.06810(28) & 2.71 \\
 96\phantom{a}&\phantom{--}0.069(37) & 0.06858(32) & 1.78 \\
128\phantom{a}&\phantom{--}0.095(39) & 0.06875(39) & 1.95 \\
\hline
\end{tabular}
\end{center}
\end{table}

\begin{table}
\caption{\sl \label{binderVil}
Fits of the Binder cumulant of the Villain model
at $\beta=0.75154$ with the ansatz~(\ref{Binderansatz}).
}
\begin{center}
\begin{tabular}{|r|l|r|c|}
\hline
$L_{min}$&\phantom{000}$C$  & $c_2$\phantom{00000}  & $\chi^2/$ d.o.f.\\
\hline
 16\phantom{a}& 1.030(19) & 0.06794(26)  & 3.32 \\
 32\phantom{a}& 1.146(37) & 0.06874(34)  & 1.12 \\
 64\phantom{a}& 1.174(72) & 0.06886(43)  & 1.35 \\
\hline
\end{tabular}
\end{center}
\end{table}

\begin{table}
\caption{\sl \label{binderASOS}
Fits of the Binder cumulant of the dual of the ASOS model 
at $\tilde \beta=0.80608$ with the ansatz~(\ref{Binderansatz}). 
}
\begin{center}
\begin{tabular}{|r|l|r|r|}
\hline
$L_{min}$&\phantom{000}$C$  & $c_2$\phantom{00000}  & $\chi^2/$ d.o.f.\\
\hline
 32\phantom{a}& --0.933(3) &   0.07197(10) &164.8\phantom{000}\\
 48\phantom{a}& --1.069(5) &   0.06936(12) &24.9\phantom{000} \\
 64\phantom{a}& --1.127(7) &   0.06844(13) & 7.17\phantom{00} \\
 96\phantom{a}& --1.180(10)&   0.06769(17) & 0.67\phantom{00} \\
128\phantom{a}& --1.184(14)&   0.06764(21) & 0.77\phantom{00} \\
\hline
\end{tabular}
\end{center}
\end{table} 

Finally, let us discuss the behaviour of the Binder cumulant at the KT-transition.
Taking e.g. the  result of the fit for the dual of the ASOS model
for $L_{min}=96$ and $\tilde \beta=0.80608$ 
one obtains that the minimum of the Binder cumulant at the transition 
is $U_{4,min}\approx 1.017006$  at $L\approx 6200$. To see the Binder cumulant
increase again, rather large lattice sizes are needed:
To get a value $1.01701$ we have to go up to $L\approx 11000$ and $1.01705$ is 
reached at about $L=38000$.  I.e. it is impossible with todays 
computing resources to see explicitly from Monte Carlo simulations that 
the Binder cumulant at the KT-transition is an increasing function of 
$L$ for sufficiently large $L$.

\section{The matching method}
\label{matchsection}
The RG-flow in the neighbourhood of the KT-transition is given, up to irrelevant
scaling fields, in terms of two coupling constants, 
$x = \pi  \beta_{SW} -2$ and the fugacity $z$ \cite{Jo77,AmGoGr80}.
Therefore, in finite size scaling we can map the KT-flow using two different 
phenomenological couplings $R_1(x(L),z(L))$ and $R_2(x(L),z(L))$.
In the ideal case, 
\begin{eqnarray}
\label{ideal1}
R_1(x,z) &=& R_1(0,0)+ c_1 x +\mbox{O}(z^2,xz) \\
\label{ideal2}
R_2(x,z) &=& R_2(0,0)+ c_2 z +\mbox{O}(x^2,xz)   \;.
\end{eqnarray} 
In the SOS representation, such pairs of phenomenological couplings can be 
easily found \cite{HaMaPi94,HaPi97}. In the XY representation, on a finite lattice
with periodic boundary conditions vortices come in pairs. Therefore
only phenomenological couplings of the type~(\ref{ideal1}) can be constructed.
\footnote{We have not studied free boundary conditions, where
also a single vortex might exist. For such boundary conditions we expect
larger power law corrections than for periodic boundary conditions.}

Since both the helicity modulus $\Upsilon$ and the second moment correlation 
length over the lattice size $\xi/L$ can be well fitted by 
eqs.~(\ref{centralheli},\ref{xiexact}) it is very hard numerically to 
observe any explicit $z$-dependence of these quantities.
Therefore it is actually fortunate that the Binder cumulant displays a
$z^2$ contribution with an apparently large amplitude.

Therefore we can at least map the RG-flow in an intermediate regime of 
lattice sizes
using $\xi/L$ or $\Upsilon$ as first phenomenological coupling and $U_4$
as second phenomenological coupling. Here, 
an intermediate regime of lattice sizes means that $L$ should be large
enough such that power law corrections due to irrelevant scaling fields 
can be ignored and on the other hand $L$ is still small 
enough such that $z^2$-contributions in the numerical data of $U_4$ are  
much larger than the statistical error.

The purpose of this mapping is to allow for an accurate determination 
of $\beta_{KT}$ from relatively small lattices.
This can be achieved by solving the following set of equations:
\begin{eqnarray}
\label{matching}
R_{1,NM}(L,\beta_{KT,NM}) & = & R_{1,SM}(\tilde b_m L,\beta_{KT,SM}) \nonumber \\
R_{2,NM}(L,\beta_{KT,NM}) & = & R_{2,SM}(\tilde b_m L,\beta_{KT,SM})\;\;,
\end{eqnarray}
where $NM$ stands for new model, where $\beta_{KT,NM}$ is not known
and $SM$ stands for solved model. In the case 
of the solved model, $\beta_{KT}$ should be at least determined numerically 
to a high precision and also the phenomenological couplings should be 
determined accurately.
The unknowns of the eqs.~(\ref{matching}) are $\beta_{KT,NM}$ of the new model
and the matching factor $\tilde b_m$. 

The right sides of eqs.~(\ref{matching}) might be given by the results of 
our simulations discussed in the previous section. These results could 
be represented by tables. To obtain values of the phenomenological couplings
for any lattice size $L_{min} \le L \le L_{max}$, where $L_{min}$ and $L_{max}$ 
are the minimal and maximal lattice size that are simulated, 
one might e.g. interpolate the values contained in the table  
in linearly $\ln(L)$.

Here instead we follow an approach that is simpler to implement. 
As we have seen in the 
previous section, our numerical data for $\xi/L$ and $U_4$ 
are well described by the ans\"atze~(\ref{xi2fits},\ref{Binderansatz}).
Therefore we suggest to use these, along with the results for the parameters, 
as right side of eqs.~(\ref{matching}).
Since in ref. \cite{HaMaPi94,HaPi97} the transition temperature is most accurately 
determined for the dual of the ASOS model, we use the data of this model
for the matching.
\newpage
In particular we take the result obtained from $L_{min}=96$ at 
\begin{eqnarray}
\label{match80608}
\tilde \beta=0.80608 &:& \nonumber \\
U_{4,ASOS}(L) &=& 1.018192 - 
  \frac{0.017922}{\ln L -1.18} + \frac{0.06769}{(\ln L -1.18)^2} \nonumber \\
 \frac{\xi_{2nd,ASOS}(L)}{L} &=& 0.7506912 + \frac{0.212430}{\ln L + 0.573}
 \;.
\end{eqnarray}
For consistency we also take $\xi/L$ at
$\tilde \beta=0.80608$ and $L_{min}=96$.
Note that in the present context, it is only important that the equation 
describes the 
data well for a certain range of lattice sizes ($96 \le L \le 1024$), where 
the matching takes place. 

To check the error due to the uncertainty of $\tilde \beta_{KT}$ one should
repeat the matching with the results obtained at
\begin{eqnarray}
\label{match80606}
\tilde \beta=0.80606 &:& \nonumber \\
U_{4,ASOS}(L) &=& 1.018192 -
\frac{0.017922}{\ln L -1.193} + \frac{0.06730}{(\ln L -1.193)^2} \nonumber \\
\frac{\xi_{2nd,ASOS}(L)}{L} &=& 0.7506912 + \frac{0.212430}{\ln L + 0.557}
\;.
\end{eqnarray}

Here we do not quote statistical errors and covariances of $C$ for 
$\xi/L$ and $c_2$ and $C$ for $U_4$. Note that applying the matching with 
eqs.~(\ref{match80608},\ref{match80606}) it makes no sense to 
produce larger statistics for the new model than  the one 
for the ASOS model here. Staying well below our present statistics, say 
$10^6$ or less measurements for the new model, the statistical error
of the coefficients in eqs.~(\ref{match80608},\ref{match80606}) can be 
safely ignored. Since we are aiming at models that are harder to simulate 
than the models discussed in the present work, this is no serious 
limitation.

\section{A first application: the $\phi^4$ model at $\lambda=2.1$}
We have tested the new method at the example of the 2-component $\phi^4$ model 
on the square lattice. 
The classical Hamiltonian is given by
\begin{equation}
 H_{\phi^4} = - \beta \sum_{x,\mu} \vec{\phi}_x \cdot \vec{\phi}_{x+\hat \mu}
 + \sum_{x} \left[\vec{\phi}_x^{\;2} + \lambda (\vec{\phi}_x^{\;2} -1)^2 \right]
\;\;,
\end{equation}
where the field variable $\vec{\phi}_x$ is a vector with two real
components. In our convention, the Boltzmann factor is given by 
$\exp(-H_{\phi^4})$. 
Since $|\vec{\phi}_x|$ is not restricted to one, the model can not be 
mapped into an SOS model. 
For our test we have chosen the value $\lambda=2.1$, which is a good
approximation of the improved value of the three dimensional model \cite{our3D}. 
In the two dimensional case this value has no particular meaning.

We have simulated the model with a mixture of the single cluster algorithm, 
the overrelaxation update and a local Metropolis update. Here the 
local Metropolis update is needed to change $|\vec{\phi}_x|$.
The single cluster algorithm, and the overrelaxation update have exactly the 
same form as in the case of the XY model. The proposal for the 
Metropolis update is given by
\begin{eqnarray}
  \phi_x^{(1)'} &=& \phi_x^{(1)} + s \left(r- \frac{1}{2} \right) \nonumber \\ 
  \phi_x^{(2)'} &=& \frac{3}{4} \beta \Phi_x^{(2)}   -  \phi_x^{(2)}  
\end{eqnarray}
with $s=3$. $r$ is a random number with a uniform distribution in the 
interval $[0,1]$. ${\Phi_x}$ is the sum over the nearest neighbours of $x$. 
The proposal for the second component of the field is constructed such that 
$H_{\phi^4}$ is only changed by little.
The proposal is accepted with the probability $A=\mbox{min}[1,\exp(-\Delta H)]$.
This Metropolis update is followed by a second update at the 
same site, where the role of the two components of the field is exchanged. 
These updates were 
performed the following sequence: One Metropolis sweep, one sweep with 
the overrelaxation algorithm, 10 single cluster updates,
followed by 5 sweeps with the overrelaxation algorithm.
After such a sequence of updates, a measurement of the observables is 
performed.

We have simulated  $L=32$ at $\beta=1.075$ and $1.08$, 
$L=64$ at $\beta=1.078$ and $1.08$, $L=128$ at $\beta=1.077$ and $\beta=1.079$
and $L=256$ at $\beta=1.0775$ and $\beta=1.0785$.  In all cases we have
performed $10^7$ measurements.
For fitting the data with the ans\"atze~(\ref{helifits},\ref{xi2fits}) 
and for the matching~(\ref{matching}), 
we linearly interpolated the values of $\Upsilon$, $\xi/L$ and $U_4$ in $\beta$. 

First we have fitted the helicity modulus to the ansatz~(\ref{helifits}), where 
$\beta_{KT}$ and the integration constant $C$ are the free parameters.
The results of these fits are summarized in table \ref{phiheli}. 
$\chi^2/$d.o.f. is quite large and the result for $\beta_{KT}$  is decreasing 
with increasing $L_{min}$. 
\begin{table}
\caption{\sl \label{phiheli}
Fitting $\Upsilon$ of the $\phi^4$ model at $\lambda=2.1$ with the analogue
of the ansatz~(\ref{helifits}).
}
\begin{center}
\begin{tabular}{|r|r|r|c|}
\hline
\multicolumn{1}{|c}{$L_{min}$} & \multicolumn{1}{|c}{$\beta_{KT}$} & 
\multicolumn{1}{|c}{$C$} & \multicolumn{1}{|c|}{$\chi^2/$d.o.f.} \\
\hline
32        & 1.07865(1)   & 0.278(2) & 75.9 \\
64        & 1.07838(2)   & 0.363(7) & 22.6 \\
128       & 1.07818(2)   & 0.447(8) &  - \\
\hline
\end{tabular}
\end{center}
\end{table} 

Next we have fitted the data for $\xi/L$ to the ansatz~(\ref{xi2fits}). 
The results are summarized in table \ref{phixi}.
The $\chi^2/$d.o.f. is already smaller than
two for $L_{min}=32$. But on the other hand also the statistical error of 
$\beta_{KT}$ obtained from  $\xi/L$  is about 6 times larger than for that 
from the helicity modulus.
In the case $\xi/L$ the estimate of $\beta_{KT}$ is an increasing function
of $L_{min}$. 
Therefore we might regard the result from $\xi/L$  and 
$L_{min}=128$ as lower bound 
and that from $L_{min}=128$ from the helicity modulus as upper bound.
Hence we quote $\beta_{KT} = 1.0779(5)$ as our
final estimate obtained from the combined analysis of $\Upsilon$ and $\xi/L$.

\begin{table}
\caption{\sl \label{phixi}
Fitting $\xi/L$ 
of the $\phi^4$ model at $\lambda=2.1$ with the analogue
of the ansatz~(\ref{xi2fits}).
}
\begin{center}
\begin{tabular}{|r|r|r|c|}
\hline
\multicolumn{1}{|c}{$L_{min}$} & \multicolumn{1}{|c}{$\beta_{KT}$} & 
\multicolumn{1}{|c}{$C$} & \multicolumn{1}{|c|}{$\chi^2/$d.o.f.} \\
\hline
32        &  1.07721(6)\phantom{0}  &  1.50(2) &1.67 \\
64        &  1.07739(12) &  1.41(6) &1.39 \\
128       &  1.07761(19) &  1.30(9) & -  \\
\hline
\end{tabular}
\end{center}
\end{table} 

Next we applied the matching method discussed in the previous section.
\begin{table}
\caption{\sl \label{match1}
Matching the $\phi^4$ model at $\lambda=2.1$ with the 
dual of the ASOS model at $\beta=0.80608$, eqs.~(\ref{match80608}).
}
\begin{center}
\begin{tabular}{|r|l|r|}
\hline
\multicolumn{1}{|c}{$L$} & \multicolumn{1}{|c}{$L_{ASOS}$} & 
\multicolumn{1}{|c|}{$\beta_{KT}$}  \\
\hline
32 & \phantom{0}55.81(12) & 1.07892(4) \\
64 & 113.6(5)  & 1.07822(3) \\
128& 234.2(1.4)& 1.07786(2) \\
256& 479.(5.)  & 1.07776(2) \\
\hline
\end{tabular}
\end{center}
\end{table} 

\begin{table}
\caption{\sl \label{match2}
Matching the $\phi^4$ model at $\lambda=2.1$ with the 
dual of the ASOS model at $\beta=0.80606$, eqs.~(\ref{match80606}).
}
\begin{center}
\begin{tabular}{|r|l|r|}
\hline
\multicolumn{1}{|c}{$L$} & \multicolumn{1}{|c}{$L_{ASOS}$} & 
\multicolumn{1}{|c|}{$\beta_{KT}$}  \\
\hline
32 &  \phantom{0}56.03(12) & 1.07898(4) \\
64 & 113.8(5)   & 1.07827(3)  \\
128& 234.0(1.4) & 1.07790(2) \\
256& 478.(5.)   & 1.07780(2) \\
\hline
\end{tabular}
\end{center}
\end{table} 

This way we obtain an estimate of $\beta_{KT}$ and the corresponding lattice 
size of the ASOS model from a single lattice size $L$. Our results for
matching with $\beta_{ASOS} = 0.80608$, eqs.(\ref{match80608}),
are given in table \ref{match1} and 
those for $\beta_{ASOS} = 0.80606$, eqs.(\ref{match80606}),
in table \ref{match2}. 
The estimates for $\beta_{KT}$ apparently converge as $L$ increases. 
Taking into account the difference between $L=128$ and $L=256$ as well as 
the difference between $\beta_{ASOS} = 0.80608$ and $\beta_{ASOS} = 0.80606$ 
we arrive at our final estimate $\beta_KT=1.0778(2)$.  This estimate is 
clearly more precise than that obtained from fitting 
$\Upsilon$ and $\xi/L$ with the ans\"atze~(\ref{helifits},\ref{xi2fits}).

\section{Summary and Conclusions}
We have studied the XY model, the Villain model and the dual of the 
absolute value solid-on-solid model on a square lattice at the 
Kosterlitz-Thouless transition.
We have focused on the behaviour of  phenomenological couplings
like the helicity modulus $\Upsilon$, 
the second moment correlation length over the lattice size $\xi/L$ and 
in particular the Binder cumulant $U_4$.

Using Monte Carlo simulations of the Gaussian model on the square lattice, 
we have determined the asymptotic value of $U_4$ and the leading 
logarithmic correction at the Kosterlitz-Thouless transition. 
See eq.~(\ref{binderleading}). 

We confronted the 
predictions~(\ref{centralheli},\ref{xiexact},\ref{binderleading}) 
for the phenomenological couplings derived from the Kosterlitz-Thouless theory
with the results of our Monte Carlo simulations. We have reached lattice
sizes of $L=4096$, $2048$ and $1024$ for the XY model, the Villain model and 
the dual of the ASOS model, respectively. We have generated several $10^6$
statistically independent configurations for each lattice size.

We find that the data for the helicity modulus $\Upsilon$ and in 
particular the 
second moment correlation length over the lattice size $\xi/L$ follow quite 
well the predictions~(\ref{centralheli},\ref{xiexact}). 

In contrast, the value of the Binder cumulant at the KT-transition is increasing 
with increasing lattice size, while eq.~(\ref{binderleading}) predicts that
the Binder cumulant should decrease for sufficiently large lattice sizes.
This discrepancy is explained by the presence of vortex pairs that have 
a distance of order $L/2$. This effect should be proportional to the fugacity
squared. Hence it should lead to a subleading correction.
Indeed adding a term $c_2/(\ln L + C)^2$ to the ansatz allows to 
fit our data for the Binder cumulant for all three models studied. The 
values for $c_2$ obtained for the three models are consistent, confirming 
the universal character of the subleading correction. 
From the results of these fits one can easily see that with the present 
computer resources it is not possible to reach sufficiently large 
lattices with high statistics to see explicitly the asymptotic behaviour 
of the Binder cumulant~(\ref{binderleading}). 

The fact that $U_4$ behaves quite differently from $\xi/L$ and $\Upsilon$
allows us to set up a matching method similar to that of 
ref. \cite{HaMaPi94,HaPi97} for solid-on-solid models. For a detailed discussion see 
section \ref{matchsection}.

As a first test, we have applied the matching method to the two component 
$\phi^4$ model on the square lattice at $\lambda=2.1$. We find that using 
moderately large lattices (up to $L=256$), we can determine the temperature
of the KT-transition quite accurately. We show that the new matching method is 
superiour to fits of $\Upsilon$ or $\xi/L$ to 
the predictions~(\ref{centralheli},\ref{xiexact}).

In the near future we plan to apply the method to thin films of the three
dimensional two component $\phi^4$ model and the dynamically diluted XY model, 
where the duality transformation to solid-on-solid models is not available.

\section{Acknowledgement}
I like to thank I. Campbell for pointing my attention to ref. \cite{wysin}.

\end{document}